\documentclass[12pt,pra,aps,amssymb,amsfonts,amsmath,tightenlines]{revtex4}

\usepackage[dvips]{graphicx}
\usepackage{dsfont, color,txfonts} 

\newcommand{\re}{\mbox{$\rm e$}}
\newcommand{\rd}{\mbox{$\rm d$}}

\begin{document}

\title{How fake news can turn against its spreader}
\author{Dorje C. Brody$^{1,2}$ and Tomooki Yuasa$^{3}$}

\affiliation{$^1$
School of Mathematics and Physics, University of Surrey, 
Guildford GU2 7XH, UK \\ 
$^2$ Department of Mathematics, Imperial College London, London SW7 2BZ, UK \\ 
$^3$ Faculty of Economics and Business Administration, Graduate School 
of Management, Tokyo Metropolitan University, Tokyo, 100-0005, Japan
}

\date{\today}

\begin{abstract}
When different information sources on a given topic are combined, they interact 
in a nontrivial manner for a rational receiver of these information sources. Suppose 
that there are two information sources, one is genuine and the other contains 
disinformation. It is shown that under the conditions that the signal-to-noise ratio of 
the genuine information source is sufficiently large, and that the noise terms in the 
two information sources are positively correlated, the effect of disinformation is 
reversed from its original intent. That is, the effect of disinformation on a receiver 
of both information sources, who is unaware of the existence of disinformation, is 
to generate an opposite interpretation. While the condition in which this 
phenomenon occurs cannot always be ensured, when it is satisfied, the effect 
provides an effective way of countering the impacts of disinformation. 
\end{abstract}

\maketitle

%%%%%%%%%%%%

\section*{Introduction} 
 
While the concept of ``fake news'' in one form of another has been around for 
centuries, the impact of disinformation on political stages worldwide has 
gained unprecedented 
attention over the past decade 
\cite{Allcott,ZZ,Kalsnes,Guarino,DGMP,AAB,Simion,FS,Yates}. To assess 
and understand the impact of disinformation on democratic 
processes, it will be useful to develop generative models that can be used to 
simulate different scenarios on how people's perceptions on a given topic 
change in accordance with the revelation of information about that topic. An 
important point to note here is the fact that any such information-driven model 
has to fall squarely 
within the framework of communication theory. The reason is simple. Whether 
a given message contains disinformation or not, communication theory is 
designed to analyse how information encoded in that message is processed 
by the receiver \cite{Wiener}. 

In communication theory, information contained in a message is decomposed 
into its signal component and noise component, while the relative magnitudes 
of the two is measured by the signal-to-noise ratio \cite{Shannon}. Exactly 
how signal and noise should be combined depends on the application, but if 
the noise is 
normally distributed, then it is natural to combine them in an additive manner, 
that is, we have a signal-plus-noise decomposition. In the simplest case, 
if the signal -- part of the message that one wishes to know -- is modelled as 
a random variable $X$, and if noise -- part of the message 
that is independent of the signal -- is modelled likewise as a random variable 
$\epsilon$, which may be normal, then transmission of noisy information 
about the signal can be modelled in the form of learning the value of the 
information variable 
\begin{eqnarray} 
\xi = \sigma X + \epsilon \, , 
\label{eq:1} 
\end{eqnarray} 
where the parameter $\sigma>0$ represents the signal-to-noise ratio of this 
message. For example, the signal $X$ may represent a set of $n$ alternatives 
labelled by $x_1$, $x_2$, $\cdots$, $x_n$, when the receiver of the message 
wishes to identify the most appropriate choice out of the $n$ alternatives, and seeks 
partial information to improve their decision. Communication theory tells us that 
the simplest form of modelling information acquisition of this kind is to make use 
of the information variable (\ref{eq:1}). This information is partial because there 
are two unknowns for the receiver, the signal $X$ and the noise $\epsilon$, and 
only one known, the information variable $\xi$. Thus acquiring the information 
$\xi$ is insufficient to determine the value of $X$, but it is sufficient to reduce 
its uncertainty \cite{Wiener}. 

Before acquiring information, that is, before the detection of $\xi$, the prior view 
of the receiver of the information about the choice $X$ is represented by the 
probabilities 
$\{p_k\}$ that $X$ takes the values $\{x_k\}$. If $p(y)$ is the density function for the 
noise variable $\epsilon$, then after the detection of $\xi$, the perception of the 
receiver is update according to the Bayes formula 
\begin{eqnarray}
p_k \to \pi_k(\xi) = \frac{p_k \, p(\xi-\sigma x_k)}{\sum_k p_k \, p(\xi-\sigma x_k)} . 
\label{eq:2} 
\end{eqnarray} 
In this way we can model the transformation $p_k\to\pi_k(\xi)$ of the perception 
of people on a given subject, based on the arrival of noisy information $\xi$. 

In this scheme of modelling the processing of noisy information using the 
framework of communication theory, there are two canonical ways, passive and 
active, to model 
disinformation in a message. The first is to modify the signal-to-noise ratio 
$\sigma$ in such a way that the receiver is unaware of the modification 
\cite{Brody5}. This can be achieved either by secretly introducing more noise, 
to reduce $\sigma$, or secretly releasing more reliable information, to increase 
$\sigma$. In either case, if the 
receiver is under the impression that the detected information is of the form 
$\xi = \sigma X + \epsilon$, while in reality it is $\xi' = \sigma' X + \epsilon$, then 
the resulting inference will be skewed. For instance, if $\sigma'<\sigma$, then 
the receiver is overconfident about the information obtained, and conversely if 
$\sigma'>\sigma$, then the receiver is under-confident about the information. 

An active form of disinformation, on the other hand, can be modelled by including 
a bias in the noise \cite{Brody1}. To this end we remark that a genuine noise by 
definition has no unknown bias. That is, the unconditional expectation of noise 
in effect vanishes so that ${\mathbb E}[\epsilon]=0$, because any known bias in 
$\epsilon$ would be discarded by the receiver anyhow. Hence if the information 
takes the form 
\begin{eqnarray} 
\xi = \sigma X + (\epsilon + f) , 
\end{eqnarray} 
where $f$ is independent of $X$ and has nonzero mean so that 
${\mathbb E}[f]\neq0$, and if in addition 
the receiver of this information is unaware of the 
existence of the noise bias $f$, then their assessment of this information is 
distorted. Specifically, the net effect is to shift the signal according to 
\begin{eqnarray} 
X \to X + f/\sigma .  
\end{eqnarray}  
Thus, if the disseminator of disinformation wishes the receiver to be misled in 
thinking that the alternatives labelled by smaller values of $x_k$ be more likely, 
then it suffices to let $f<0$. Conversely, a positive $f$ leads to an 
assessment whereby the receiver thinks those alternatives labelled by larger 
values of $x_k$ more likely. 

\begin{figure}[t!]
\centerline{
\includegraphics[width=1.0\textwidth]{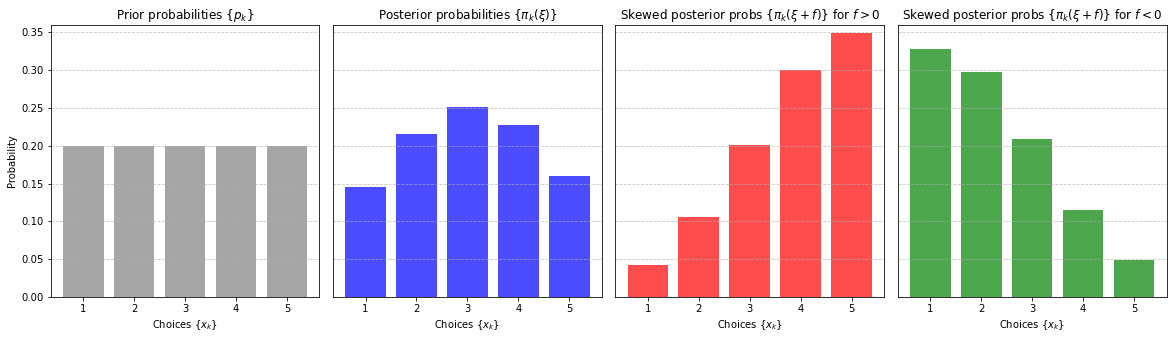}
}
\caption{\footnotesize{
\textit{Perceptions skewed by disinformation}. 
On the left panel the prior probabilities $\{p_k\}$, here taken to be uniform, for 
the random variable $X$ representing five possible alternatives, are plotted in 
the form of a histogram. Having sampled the value of $\xi=\sigma X+\epsilon$, 
where the signal-to-noise ratio is taken to be $\sigma=0.5$ and $\epsilon$ is a 
zero-mean normal variable with standard deviation $1$, the prior transforms 
into posterior $\{\pi_k(\xi)\}$ (the second panel). However, if the message 
$\xi$ is contaminated with disinformation with $f>0$, whose existence is 
unknown to the receiver, then the posterior view is skewed to the right 
(the third panel); whereas if $f<0$ then it is skewed to the left. 
}}
\label{fig1}
\end{figure}

In more specific terms, because the receiver by assumption is unaware of the 
existence of $f$, their assessment is reflected in the view modelled by 
$\{\pi_k(\xi)\}$. 
However, in reality the value of $\xi$ has been shifted by $f$, implying that their 
assessment is skewed into a modified view $\{\pi_k(\xi+f)\}$. In Figure~\ref{fig1} 
we sketch examples of how the effect of disinformation manifests itself in such 
an assessment. Starting 
with a uniform prior over five alternatives, a detection of the message $\xi$ 
turns this into the corresponding posterior view, but this is skewed towards 
one direction or the other, depending on whether $f$ is positive or negative. 

We see therefore that we arrive at a simple modelling framework that captures 
people's responses to disinformation, entirely within the framework of elementary 
communication theory. An important point to keep in mind is that the receiver is 
unaware of the existence of the noise bias $f$. This is in keeping with the fact 
that if a receiver of a message knows that it contains disinformation, that is, if the 
receiver knows the existence and the value of $f$, then that part of the message 
will be discarded, if the objective is to make the most informed choice. 

This active way of modelling disinformation was first envisaged in \cite{Brody1} 
and it has been applied to investigate the impact of disinformation in electoral 
competitions \cite{Brody2,Brody3}. The purpose of the present paper is to 
investigate the effect of combining different information sources, when one, or 
some of these sources contain disinformation. Specifically, the present paper is 
organised as follows. We first consider how two information sources on a given 
topic combine and interact. The result shows that depending on the relative 
strengths of the signal-to-noise ratios, the intended impact of a given information 
source can generate an opposite effect to a receiver who consumes both 
information sources. This means, in particular, that, should there be any 
disinformation contained in that information source, 
it will backfire. That is, fake news will 
turn against its spreader in that scenario. We then consider a time series 
generalisation of the model to capture circumstances involving a continuous 
arrival of information, and illustrate the effect in that setup. We then examine 
another phenomenon, not related to the issue on how information sources are 
combined, but nonetheless illustrates circumstances in which an excessive 
dissemination of disinformation can result in an unintended outcome for the 
disseminator of the disinformaton. We 
conclude the paper with a brief discussion on broader issues.

\section*{Combining different information sources} 

In general there can be several information sources for a given topic. To model 
this situation, let us consider first the case in which there are two information 
sources: 
\begin{eqnarray} 
\xi_1 = \sigma_1 \, X + \epsilon_1 \quad {\rm and} \quad 
\xi_2 = \sigma_2 \, X + \epsilon_2 \, . 
\label{eq:5} 
\end{eqnarray} 
For now we assume that these information sources are free of disinformation. For 
definiteness we assume that $\epsilon_1$ and $\epsilon_2$ are both zero-mean 
normal random variables with the same standard deviation, having the correlation 
$\rho$. Note that the assumption ${\mathbb E}[\epsilon_1^2]=
{\mathbb E}[\epsilon_2^2]$ here on the standard deviations of the two noise 
terms yields no loss of generality because if the 
standard deviations were different, then we can make them the same by 
rescaling the signal-to-noise ratio. 

The idea represented in (\ref{eq:5}) 
is that the receiver gets two messages on the same topic, 
modelled by $\xi_1$ and $\xi_2$. We shall now show that the aggregate of these 
messages can be expressed in the form of a single message of the form 
(\ref{eq:1}). 

We begin by remarking that because the correlation of $\epsilon_1$ and 
$\epsilon_2$ is $\rho$, there is a zero-mean normal variable ${\bar\epsilon}$, 
independent of $\epsilon_1$, such that we can write 
\begin{eqnarray}
\epsilon_2 = \rho \, \epsilon_1 + \sqrt{1-\rho^2} \, {\bar\epsilon}  \, . 
\end{eqnarray}
Analogously, we can define an auxiliary information variable ${\bar\xi}$ by the 
relation 
\begin{eqnarray}
\xi_2 = \rho \, \xi_1 + \sqrt{1-\rho^2} \, {\bar\xi}  \, . 
\label{eq:7} 
\end{eqnarray}
By a rearrangement of terms we thus have 
\begin{eqnarray}
{\bar\xi}  = \frac{\sigma_2-\rho\sigma_1}{\sqrt{1-\rho^2}} \, X + 
\frac{\epsilon_2-\rho\epsilon_1}{\sqrt{1-\rho^2}} = {\bar\sigma} \, X + 
{\bar\epsilon} ,
\end{eqnarray}
where ${\bar\sigma}$ is defined to be the coefficient of $X$ in the middle 
expression. It is evident from (\ref{eq:7}) that the specification of the two messages 
$\xi_1$ and $\xi_2$ is equivalent to that of $\xi_1$ and ${\bar\xi}$, however, the 
latter two messages have independent noise terms $\epsilon_1$ and 
${\bar\epsilon}$. If we define $\delta$ according to 
\begin{eqnarray}
\delta  = \frac{\xi_1}{\sigma_1} - \frac{{\bar\xi}}{{\bar\sigma}} = 
\frac{\epsilon_1}{\sigma_1} - \frac{{\bar\epsilon}}{{\bar\sigma}} , 
\end{eqnarray}
then $\delta$ is independent of $X$, thus represents pure noise. We now 
construct another zero-mean normal noise variable $\epsilon$ with the 
properties that it is independent of $\delta$ and that it has the same standard 
deviation as $\epsilon_1$ and $\epsilon_2$. Writing $\epsilon=\alpha\epsilon_1 
+ \beta \epsilon_2$, the independence condition ${\mathbb E}[\epsilon\delta]=0$ 
shows that $\alpha=\lambda(\sigma_1-\rho\sigma_2)$ and $\beta= \lambda 
(\sigma_2-\rho\sigma_1)$ for some $\lambda$. The variance condition 
${\mathbb E}[\epsilon^2]={\mathbb E}[\epsilon_1^2]$ then fixes $\lambda$ to 
give 
\begin{eqnarray}
\epsilon = \sqrt{\frac{1-\rho^2}{\sigma_1^2-2\rho\sigma_1\sigma_2+
\sigma_2^{2}}} \left[ \frac{\sigma_1-\rho\sigma_2}{1-\rho^2}\, \epsilon_1 + 
\frac{\sigma_2-\rho\sigma_1}{1-\rho^2} \, \epsilon_2 \right]\, ,
\label{eq:10} 
\end{eqnarray}
which, recall, is independent of $\delta$. 

To proceed it will be useful to define a parameter $\sigma$ according to 
\begin{eqnarray}
\sigma = \sqrt{ 
\frac{\sigma_1^2-2\rho\sigma_1\sigma_2+\sigma_2^2}{1-\rho^2} } \, . 
\label{eq:11}
\end{eqnarray}
In line with (\ref{eq:10}), and using (\ref{eq:11}), we now define 
a new message $\xi$ by setting 
\begin{eqnarray} 
\xi = \frac{\sigma_1-\rho\sigma_2}{\sigma(1-\rho^2)}\, \xi_1 + 
\frac{\sigma_2-\rho\sigma_1}{\sigma(1-\rho^2)} \, \xi_2 \, . 
\label{eq:12} 
\end{eqnarray} 
Then a direct substitution shows that $\xi$ can be expressed alternatively in 
the form (\ref{eq:1}), 
where $\sigma$ is defined by (\ref{eq:11}) and $\epsilon$ is defined by 
(\ref{eq:10}). Importantly, both $X$ and $\epsilon$ are independent of 
$\delta$. Further, the specification of the original pair of messages $\xi_1$ 
and $\xi_2$ is \textit{equivalent} to that of the pair $\xi$ and $\delta$, 
because there is an invertible linear transformation between these message 
pairs: 
\begin{eqnarray} 
\left( \begin{array}{c} \xi \\ \delta \end{array} \right) 
 = 
\left( \begin{array}{cc}  
\frac{\sigma_1-\rho\sigma_2}{\sigma(1-\rho^2)} & 
\frac{\sigma_2-\rho\sigma_1}{\sigma(1-\rho^2)} \\  
\frac{\sigma_2}{\sigma_1(\sigma_2-\rho\sigma_1)} & 
-\frac{1}{\sigma_2-\rho\sigma_1} \end{array} \right) 
\left( \begin{array}{c} \xi_1 \\ \xi_2 \end{array} \right) \, . 
\label{eq:12.5} 
\end{eqnarray} 
However, because $\delta$ is independent of $X$ and $\epsilon$, knowledge 
of $\delta$ makes no contribution towards the assessment of the choice 
represented by $X$. In other words, without loss we can discard $\delta$. It 
follows that the two messages of (\ref{eq:5}) can be modelled in the form of 
a single message of the same form (\ref{eq:1}). In fact the same conclusion 
follows even if there are more than two information sources; they will combine 
to be modelled by a single effective information source (\ref{eq:1}). Hence we 
can interpret (\ref{eq:1}) as representing the aggregate of all messages on 
the topic modelled by $X$. 

With the result (\ref{eq:12}) at hand we are in the position to assess the 
impact of disinformation when messages are combined. Specifically, 
suppose that the first message contained disinformation $f_1$ and the second 
$f_2$. Then from (\ref{eq:12}) we deduce that when the two messages are 
combined, the net effect is to generate a shift 
\begin{eqnarray} 
\xi \to \xi + \frac{\sigma_1-\rho\sigma_2}{\sigma(1-\rho^2)}\, f_1 + 
\frac{\sigma_2-\rho\sigma_1}{\sigma(1-\rho^2)} \, f_2 \, . 
\label{eq:13} 
\end{eqnarray} 
If the correlation $\rho$ of the two noise terms $\epsilon_1$ and $\epsilon_2$ 
is negative, then there are no surprises here, whereas if $\rho>0$, then there 
can be an unexpected effect of disinformation. It suffices to consider the 
contribution of, say, $f_1$. From the form $\xi_1 = \sigma_1 (X+f_1/\sigma_1) + 
\epsilon_1$ of the message we see that a positive $f_1$ is intended to mislead 
the receiver in 
thinking that the alternatives $x_k$ with larger values of $k$ more likely than 
what it actually is. Conversely, if $f_1<0$ then the intention is to mislead the 
receiver in thinking that those alternatives with smaller values of $k$ are more 
plausible. This is the clear effect of $f_1$ on $\xi_1$, and similarly for $f_2$ on 
$\xi_2$. However, when two messages are combined, the effect of $f_1$ 
and $f_2$ on the aggregate information $\xi$ is to generate a shift 
\begin{eqnarray}
X \to X + \frac{\sigma_1-\rho\sigma_2}{\sigma^2(1-\rho^2)}\, f_1 
+ \frac{\sigma_2-\rho\sigma_1}{\sigma^2(1-\rho^2)} \, f_2 
\label{eq:14} 
\end{eqnarray}
of the signal. 
Observe that when $\rho>0$, it is possible that one of the coefficients of $f_1$ 
and $f_2$ be negative (they can be both positive but they cannot both be 
negative). If that were the case, then the unintended consequence is that the 
disinformation will generate an opposite effect. 

As an example, suppose that $\xi_1$ represents a genuine, albeit noisy, 
information, while $\xi_2$ contains a deliberate disinformation $f_2$. In this case, 
if $\sigma_2<\rho\sigma_1$, then any such disinformation contained in $\xi_2$ 
will generate a contrary effect on the receiver of both information sources 
$\xi_1$ and $\xi_2$. We can think, for instance, of a democratic process, such 
as an election, for which the signal $X$ labels different candidates. In the 
simplest electoral model setup, the numerical values of the gaps $x_k-x_l$ can 
be interpreted as representing the relative differences of the positions of the 
candidates (or their political parties) on the political spectrum \cite{Brody4}. 
(While more detailed structural models taking into account the various factors 
that affect voter concerns as well as voter demographics are available 
\cite{Brody1}, the simplest reduced-form model considered here is sufficient to 
capture qualitative features of the information-based model we wish to highlight 
here.) 
Under the convention that $x_n=\max(x_k)$ represents the position that is the 
farthest to the right and similarly $x_1=\min(x_k)$ representing the farthest to the 
left, a far-right political party naturally wishes the electorates to make the choice 
$X=x_n$, and similarly for the far-left wishing the choice $X=x_1$. 
A recent study \cite{Tornberg} suggests that political parties at the right-end of 
the spectrum are significantly more likely to disseminate disinformation. In such 
a context we can think of $\xi_1$ representing information emanating from 
centre/left-leaning parties, and $\xi_2$ from the right-leaning parties. Any 
disinformation contained in $\xi_2$ will thus have the property that $f_2>0$. 

When the condition $\sigma_2<\rho\sigma_1$ is met, the impact of such 
disinformation on the electorates who consume both information sources 
(and who is unaware of the existence of disinformation) is that the stronger 
the disinformation is, the less likely it is that they will vote for right-leaning 
candidates -- in spite of the fact that they are unaware of the existence of 
disinformation. As a strategy to counter the impact of disinformation, 
therefore, it suffices to increase the value of $\sigma_1$, provided that 
$\rho>0$. Intuitively speaking, what this means is that rather than investing 
energy into denying disinformation, it is more effective simply to release 
clear and strong message. 

Needless to add, such a strategy does not affect those in the so-called 
disinformation bubble -- for instance those who only consume messages 
represented by $\xi_2$. Also, the positive correlation condition $\rho>0$ 
is in general not enforceable by a single information source. Recall that 
$\rho$ is the correlation between $\epsilon_1$ and $\epsilon_2$; the two 
noise terms, where by noise we mean rumours, speculations, ambiguities, 
misinterpretations, and the like. The nature of noise is typically dependent 
on the communication channel -- for example, the media or the internet -- 
and therefore a single information source cannot unanimously fix the value 
of $\rho$. Nevertheless, when the condition $\sigma_2<\rho\sigma_1$ is met, 
the strategy suggested here will be highly effective to counter the impact of 
disinformation, at least based on what communication theory tells us. 

Before we proceed further, it will be useful to discuss the intuition behind the 
surprising 
fact that when $\sigma_2<\rho\sigma_1$, any disinformation $f_2$ will 
generate an unintended effect for the disseminator of disinformation. While the 
mathematical analysis leading to the result (\ref{eq:14}) is very simple, the effect 
is nevertheless counterintuitive. To understand this phenomenon, recall that the 
two messages take the form $\xi_1=\sigma_1 X+\epsilon_1$ and 
$\xi_2=\sigma_2 X+\epsilon_2+f_2$, where the existence of the term $f_2$ is 
unknown to the receiver. Now suppose that the strength of the message 
$\xi_1$ is sufficiently large when compared to that of $\xi_2$, i.e. 
$\sigma_1>\sigma_2/\rho$. Suppose also that $f_2$ takes on a large positive 
value. The intention therefore is to mislead the receiver in thinking that $X$ 
should take a large value, i.e. to enhance the chance of the receiver selecting 
the alternatives labelled by larger values of $x_k$. This means that $\xi_2$ 
will take on a large value, while $\xi_1$ will not. However, the signal strength 
of $\xi_1$ is strong, so logically, without the knowledge of the existence of $f_2$, 
the fact that $\xi_2$ is large implies to the receiver that 
the associated noise $\epsilon_2$ must have taken on a large value. Because 
$\epsilon_1$ is positively correlated to $\epsilon_2$, it is likely therefore that 
$\epsilon_1$ must have taken a large value, but $\xi_1$ did not take a large 
value. It follows, therefore, that $X$ must have take on a small value, not large -- 
contrary to the intention of the disinformation. This 
is the logic underpinning the phenomenon uncovered here. 

\section*{Information flow in continuous time} 

In the previous section we examined how a rational receiver would respond to 
a message, or a set of messages, at a single moment in time. 
In reality people consume information regularly, so we wish 
to extend the `single-shot' information model (\ref{eq:1}) into a form of a time 
series. In general, the quantity of interest, or the signal, the noise, and the 
signal-to-noise ratio can all vary in time. However, here we consider 
the case in which the signal remains to be modelled by a fixed random variable 
$X$ but noise takes the form of a time series $\{\epsilon_t\}$. The rationale for 
considering a fixed signal is that when considering the impact of disinformation 
in an electoral competition, which is our primary interest here, it is relatively rare 
that a candidate is replaced in the middle of an election cycle. If the 
signal-to-noise ratio is given by a time series $\{\sigma_t\}$, then the time series 
representing the arrival of information now takes the form 
\begin{eqnarray} 
\xi_t = X \int_0^t \sigma_s \, \rd s + \epsilon_t \, . 
\label{eq:15} 
\end{eqnarray} 
This can be interpreted as the time-series version of the simple model (\ref{eq:1}), 
when the signal is fixed in time. 
Because we are working in the context of Gaussian-noise models, it suffices to 
assume that $\{\epsilon_t\}$ is a standard Brownian motion, which is a Gaussian 
process with mean zero and standard deviation $\sqrt{t}$, having statistically 
independent increments. 

As before, the signal $X$ represents the choices to be made by the receiver of 
the information, wherein different alternatives will be modelled by the numbers 
$\{x_k\}$. The \textit{a priori} probability that $X=x_k$, that is, the probability that 
the $k$th alternative being selected, is given by $p_k$. For simplicity of 
exposition, let us for now assume that the signal-to-noise ratio, or the information 
flow rate, is constant in time so that $\sigma_t=\sigma$. Then the time series 
representing the message takes on a simple form $\xi_t = \sigma X t + \epsilon_t$. 
Having detected the message up to time $t$, a rational receiver of the message 
will update their views on the different alternatives in accordance with the Bayes 
formula 
\begin{eqnarray}
p_k \to \pi_{k}(\xi_t,t) = \frac{p_k \, \re^{\sigma x_k \xi_t - \frac{1}{2}\sigma^2 
x_k^2 t}}{\sum_k p_k \, \re^{\sigma x_k \xi_t - \frac{1}{2}\sigma^2 
x_k^2 t}} ,
\end{eqnarray} 
which, in the context of signal processing is also known as the Wonham 
filer \cite{wonham}. 
The fact that the posterior probabilities are functions of $\xi_t$ but not on its 
history is a consequence of having a time-independent signal-to-noise ratio, 
making $\xi_t = \sigma X t + \epsilon_t$ a Markov process. For a 
time-dependent signal-to-noise ratio, on the other hand, the posterior 
probabilities will be dependent on the entire history of $\{\xi_t\}$ up to time $t$ 
\cite{Brody2}. 

We examine now the case in which there are two information-providing time 
series 
\begin{eqnarray} 
\xi^1_t = \sigma_1 X t + \epsilon^1_t \quad {\rm and} \quad 
\xi^2_t = \sigma_2 X t + \epsilon^2_t \, ,
\label{eq:17} 
\end{eqnarray} 
where the Brownian motions $\{\epsilon^1_t,\epsilon^2_t\}$ are independent of 
the signal $X$ and have the correlation $\rho$. Then the analysis leading to 
equation (\ref{eq:12}) remains valid and we find that the information contents of 
the two messages in (\ref{eq:17}) can be combined into a single information 
process $\xi_t = \sigma X t + \epsilon_t$, where $\sigma$ is given by (\ref{eq:11}) 
and $\epsilon_t$ is given by (\ref{eq:10}), with $\epsilon_1$ and $\epsilon_2$ 
replaced by $\epsilon^1_t$ and $\epsilon^2_t$. In particular, if the two 
messages were to contain disinformation terms $f^1_t$ and $f^2_t$, for instance 
noise-biasing drifts of the form $f^1_t=\mu_1t$ and $f^2_t=\mu_2t$, then this 
will result in shifting the combined information according to 
\begin{eqnarray} 
\xi \to \xi + \frac{\sigma_1-\rho\sigma_2}{\sigma(1-\rho^2)}\, \mu_1 t + 
\frac{\sigma_2-\rho\sigma_1}{\sigma(1-\rho^2)} \, \mu_2 t \, . 
\label{eq:18} 
\end{eqnarray} 
Hence as in the previous example, if $\sigma_1<\rho\sigma_2$ then any 
intentional disinformation in the message $\xi^1_t$ will work against its spreader, 
and similarly if $\sigma_2<\rho\sigma_1$ then any intentional disinformation 
in the message $\xi^2_t$ will backfire. 

\begin{figure}[t!]
\centerline{
\includegraphics[width=1.0\textwidth]{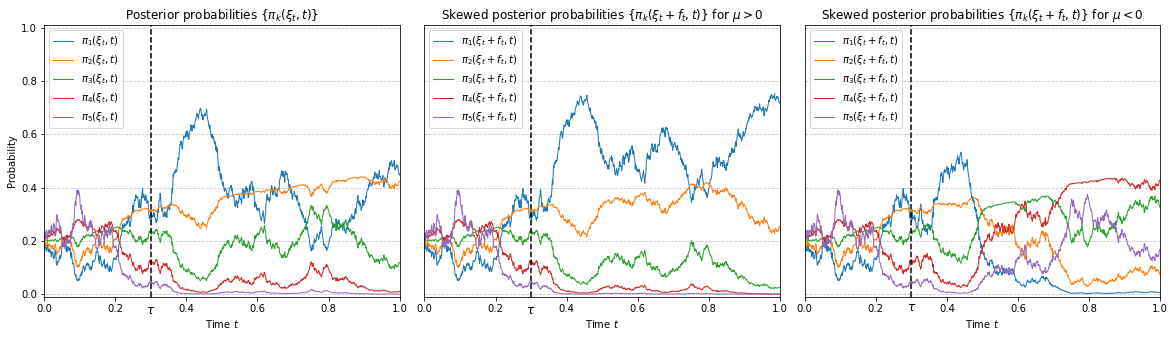}
}
\caption{\footnotesize{
\textit{Sample paths showing the reversal of the impact of disinformation}. 
On the left panel the posterior probabilities $\{\pi_k\}$ are plotted when $X$ 
represents five alternatives, in the absence of disinformation. When the 
second information source contains disinformation of the form $f^2_t = 
\mu (t-\tau) {\mathds 1}\{t>\tau\}$ that is released at time $\tau$, the 
\textit{intension}  
is to skew the preference of the receiver towards $\pi_5$ if $\mu>0$. Similarly, 
if $\mu<0$ then the intension is to skew the probabilities towards $\pi_1$. 
However, when $\sigma_2<\rho\sigma_1$, the impact of disinformation is 
reversed, as shown in the middle panel for $\mu>0$ depicting how the 
histories on the left panel would have been affected by the presence of 
disinformation, enhancing $\pi_1$ rather than $\pi_5$. Similarly, the right 
panel shows the effect for $\mu<0$, enhancing $\pi_5$ rather than $\pi_1$. 
}}
\label{fig2}
\end{figure}

So far we have considered the case involving two major information sources. 
When there are multiple information sources, their combined effect on a 
receiver who consumes all of them is more subtle. Specifically, suppose that 
there are $n$ information sources on $X$ of the form (\ref{eq:17}). Then writing 
$\rho_{ij}$ for the correlation of the noise terms $\epsilon^i_t$ and 
$\epsilon^j_t$, the aggregate of these $n$ information sources is equivalent to 
the consumption of a single information 
\begin{eqnarray}
\xi_t = \frac{1}{\sigma} \sum_{i,j} \sigma_i \,\rho^{-1}_{ij} \,\xi^j_t , 
\end{eqnarray} 
where $\rho^{-1}_{ij}$ denotes the elements of the inverse correlation matrix and 
\begin{eqnarray}
\sigma^2 = \sum_{i,j} \sigma_i  \, \rho^{-1}_{ij} \,\sigma_j \, . 
\end{eqnarray}
Therefore, with the information on noise correlation it is possible to identify the 
impacts of various disinformation terms. However, in practice the estimation of 
the noise correlation in a multi-media context will be challenging. Nevertheless 
it is entirely possible that for a fixed $j$ the sum $\sum_i \sigma_i \, \rho^{-1}_{ij}$ 
becomes negative, and in this case any disinformation that might be contained 
in the information source $\xi^j_t$ with an intent to manipulate people in one 
direction will result in the opposite direction.

\section*{The overshoot effect} 

We now turn to another unintended consequence that might arise as a result of 
disseminating disinformation, quite distinct from the reversal effect discussed in 
the foregoing discussion. To this end we note that the numerical values $\{x_k\}$ 
that the random variable $X$ takes represent, in the context of an electoral 
competition, the policy positions of the different candidates on an issue modelled 
by $X$. In the simplest case considered here of having just one issue, we may 
think of them as representing positions of the candidates, or their political parties, 
on the political spectrum. Without loss we may assume that $x_1<x_2<\cdots<
x_n$, where $x_n$ represents the farthest position to the right and $x_1$ the 
furthest to the left. 

The overshoot effect discussed below 
can occur to candidates labelled by $x_2$, $x_3$, $\cdots$, 
$x_{n-1}$, that is, those who are not at the end of the spectrum, and it can be 
explained through the following example. Suppose that the majority of the 
electorates prefer central positions, and that the political party labelled, say, by 
$x_{n-1}$ wishes to influence the electorates towards further right by means of 
a dissemination of disinformation. Then for those who consume the information 
disseminated by that political party, their preferences will shift towards right, but 
too much disinformation will move them too far towards $x_n$, bypassing 
$x_{n-1}$. This is the overshoot effect. 

At a purely anecdotal level, one can think of the example of the UK Conservative 
party, the oldest modern political party in the world, over the past five years or so. 
During this period, in particular leading up to the 2024 general election, 
political messages from the Conservative government were very much focused 
on an exaggerated statements about the dangers of migration into the UK. 
Indeed, ``stop the boat'' was a catch phrase regularly used like a prayer by the last 
Conservative prime minister during this period, in reference to the boats carrying 
undocumented migrants heading towards the UK. The reaction of the electorates, 
however, is to turn to a newly formed Reform UK party, who placed themselves 
at the far end of the spectrum in relation to their stance against migration. In 
other words, the messaging probably did have some effects, except that they 
were unintended ones. 

The phenomenon can be understood quite intuitively, but communication theory 
provides us with both qualitative and quantitative understanding. In essence, the 
phenomenon is related to the fact that while $\max_{\xi}\pi_1(\xi) = 
\max_{\xi}\pi_n(\xi) = 1$, $\max_{\xi}\pi_k(\xi)<1$ for $k\neq 1,n$. That is, while 
the posterior probabilities $\pi_1(\xi)$ and $\pi_n(\xi)$ of the extreme ends of 
the spectrum are unbounded in the range $[0,1]$, those of the intermediate 
probabilities are bounded by values strictly less than one. To see this, consider 
the example whereby the noise density is normal with mean zero and variance 
one. Then we have from the Bayes formula (\ref{eq:2}) that 
\begin{eqnarray}
\pi_k(\xi) = \frac{p_k \, \re^{\sigma x_k \xi - \frac{1}{2}\sigma^2x_k^2}}
{\sum_k p_k \, \re^{\sigma x_k \xi - \frac{1}{2}\sigma^2x_k^2}} . 
\label{eq:21} 
\end{eqnarray} 
It follows at once that 
\begin{eqnarray}
\frac{\partial}{\partial \xi}\, \pi_k  = \sigma \pi_k \left( x_k - \sum_k x_k \pi_k \right) . 
\label{eq:22} 
\end{eqnarray} 
Because the expectation ${\mathbb E}[X|\xi]=\sum_k x_k \pi_k$ appearing here 
is monotonic in $\xi$, it follows that the functions $\pi_k(\xi)$, $k=2,3,\ldots,n-1$, 
have single peaks at the values of $\xi$ for which ${\mathbb E}[X|\xi]=x_k$. It 
follows that if the value of $\xi$ is too large, perhaps due to an excessive 
dissemination of disinformation by the candidate labelled, say, by $x_{n-1}$, 
then the likelihood $\pi_{n-1}(\xi)$ will start to decrease, to be overtaken 
by the likelihood $\pi_n(\xi)$.

\section*{Discussion} 

In summary, we have presented two unintended consequences of disseminating 
disinformation --- unintended from the viewpoint of the disseminator. These are 
concerned with how different information sources combine, and how excessive 
disinformation can push people away into further extreme positions. 
Although conditions for which back reactions of disinformation occur cannot 
always be ensured, when they are met, the theory outlined here on `information 
fusion' provides effective measures to counter the impacts of disinformation.  

In general, countering the impacts of disinformation is challenging. Fact checkers 
are useful inasmuch as deciding whether any particular information is genuine or 
not, but such retrospective analysis have limited effect on countering the impacts. 
On the other hand, it has been shown that with the statistical information about 
the disinformation term $f$, a receiver can eradicate the overall majority of the 
impact of disinformation \cite{Brody1}. That is, even if one does not know whether 
a given information is true or false, the knowledge of the statistical distribution of 
disinformation alone is sufficient to eradicate the intended impact of disinformation. 
In terms of modelling, this corresponds to the situation in which the receiver of a 
message $\xi = 
\sigma X + \epsilon + f$ does not know the value of $f$ but is aware of its 
existence and its distribution. What this means in practice is that data on the 
statistics of disinformation --- evidently available to fact checkers but are rarely 
made accessible to the public --- are more valuable as a tool to counter the 
overall impacts of disinformation than the outcomes of individual fact checking 
analysis. 

Recent political shifts in the US and elsewhere make the availability of such 
data less likely. Against this background, any innovation to counter the impacts 
of disinformation seems valuable towards defending democratic processes, 
even if their scopes are limited.

\vspace{0.45cm} 
\noindent 
{\bf Acknowledgements}.
%The authors thank Bernhard Meister for stimulating discussion. 
%DCB acknowledges support from the EPSRC 
%(EP/X019926) and the John Templeton Foundation (62210). The opinions 
%expressed in this publication are those of the authors and do not necessarily 
%reflect the views of the John Templeton Foundation. 
TY is supported by JSPS KAKENHI (22K13965).


\begin{thebibliography}{}

\bibitem{Allcott} 
Allcott,~H. \& Gentzkow,~M. (2017) 
Social media and fake news in the 2016 election.
{\em Journal of Economic Perspectives}, \textbf{31}, 211--236. 
\url{doi.org/10.1257/jep.31.2.211}

\bibitem{Kalsnes} 
Kalsnes,~B. (2018) 
Fake news. 
Oxford Research Encyclopedia of Communication, 2018
\url{doi.org/10.1093/acrefore/9780190228613.013.809}

\bibitem{ZZ}
Zhou,~X. \& Zafarani,~R. (2020) 
A survey of fake news: Fundamental theories, detection methods, and 
opportunities. 
{\em ACM Computing Surveys} \textbf{53}, 109. 

\bibitem{Guarino}
Guarino,~S., Trino,~N., Celestini,~A., Chessa,~A. \& Riotta,~G. (2020) 
Characterizing networks of propaganda on twitter: a case study. 
{\em Applied Network Science} \textbf{5}, 59. 
\url{doi.org/10.1007/s41109-020-00286-y}

\bibitem{DGMP} 
D'Ambrosio,~R., Giordano,~G., Mottola,~S. \& Paternoster,~B. (2021) 
Stiffness analysis to predict the spread out of fake information. 
{\em Future Internet} \textbf{2021}, 222.  
\url{doi.org/10.3390/fi13090222}

\bibitem{AAB}  
Aïmeur,~E., Amri,~S. \& Brassard,~G. (2023) 
Fake news, disinformation and misinformation in social media: a review. 
{\em Social Network Analysis and Mining} \textbf{13}, 30. 
\url{doi.org/10.1007/s13278-023-01028-5}

\bibitem{Simion} 
Simion,~M. (2023) 
Knowledge and disinformation. 
{\em Episteme} \textbf{XX}, 1--12. 
\url{doi.org/10.1017/epi.2023.25}

\bibitem{FS} 
Farkas,~J. \& Schou,~J. (2023) 
{\em Post-truth, Fake News and Democracy}, 2nd ed. 
New York: Routledge. 

\bibitem{Yates} 
Sontag,~A., Rogers,~T. \& Yates,~C.~A. (2024) 
Dynamics of information networks. 
{\em Journal of Applied Probability} \textbf{61}, 1029-1039. 
\url{10.1017/jpr.2023.91}

\bibitem{Wiener} 
Wiener,~N.  (1948) 
{\em Cybernetics, or Control and Communication in the Animal and
the Machine}. 
Boston, MA: The Technology Press of the MIT.

\bibitem{Shannon} 
Shannon,~C.~E. \& Weaver,~W.  (1949) 
{\em The Mathematical Theory of Communication}. 
Chicago: University of Illinois Press.

\bibitem{Brody5} 
Brody,~D.~C. \& Law,~Y.~T. (2015) 
Pricing of defaultable bonds with random information flow. 
{\em Applied Mathematical Finance} \textbf{22}, 399-420. 
\url{doi.org/10.1080/1350486X.2015.1050151}

\bibitem{Brody1} 
Brody,~D.C. \& Meier,~D.~M. (2022) 
Mathematical models for fake news. In 
{\em Financial Informatics: An Information-Based Approach to Asset Pricing}, 
D.~C.~Brody, L.~P.~Hughston, A.~Macrina (eds) (Singapore: World Scientific). 
(First appeared in 2018 in \url{https://arxiv.org/abs/1809.00964})  

\bibitem{Brody2} 
Brody,~D.C. (2019) 
Modelling election dynamics and the impact of disinformation. 
{\em Information Geometry} \textbf{2}, 209-230. 
\url{doi.org/10.1007/s41884-019-00021-2} 

\bibitem{Brody3} 
Brody,~D.~C. (2022) 
Noise, fake news, and tenacious Bayesians. 
{\em Frontiers in Psychology} \textbf{13}, 797904. 
\url{doi.org/10.3389/fpsyg.2022.797904}

\bibitem{Brody4} 
Brody,~D.C. \& Yuasa,~T. (2023) 
Three candidate election strategy.   
{\it Royal Society Open Science} \textbf{10}, 230584. 
\url{doi.org/10.1098/rsos.230584}

\bibitem{Tornberg} 
T\"ornberg,~P. and Chueri,~J. (2025) 
When do parties lie? Misinformation and radical-right populism across 26 
countries. 
{\em The International Journal of Press/Politics} \textbf{30} (to appear). 
\url{doi.org/10.1177/19401612241311886} 

\bibitem{wonham}
Wonham,~W.~M. (1965) 
Some applications of stochastic differential equations to optimal nonlinear filtering.
\textit{Journal of the Society for Industrial and Applied Mathematics} 
A\textbf{2}, 347-369.
\url{doi.org/10.1137/030202}

\end{thebibliography}
\end{document}